\documentclass[seceq]{ptptex}

\usepackage[dvips]{graphicx,color}
\usepackage{multirow}
\usepackage{bm}
\usepackage{wrapft}

\def\la{\mathrel{\mathpalette\fun <}}
\def\ga{\mathrel{\mathpalette\fun >}}
\def\fun#1#2{\lower3.6pt\vbox{\baselineskip0pt\lineskip.9pt
\ialign{$\mathsurround=0pt#1\hfil##\hfil$\crcr#2\crcr\sim\crcr}}}

\def\g{\gamma}
\def\d{\delta}
\def\ve{\varepsilon}

\title{
Description of Four-Body Breakup Reaction with the Method of
Continuum-Discretized Coupled-Channels%
}

\author{
Tomoaki \textsc{Egami},$^1$ Takuma \textsc{Matsumoto},$^2$
Kazuyuki \textsc{Ogata}$^1$ and Masanobu \textsc{Yahiro}$^1$%
}

\inst{
$^1$Department of Physics, Kyushu University, Fukuoka 812-8581, Japan\\
$^2$RIKEN Nishina Center, Wako 351-0198, Japan%
}

\abst{
 We present a method of smoothing discrete breakup $S$-matrix elements
 calculated by the method of continuum-discretized coupled-channels
 (CDCC). This smoothing method makes it possible to apply CDCC to
 four-body breakup reactions. The reliability of the smoothing method is
 confirmed for two cases, $^{58}$Ni($d$, $p n$) at 80~MeV
 and the E1 transition of $^6$He. We apply CDCC with the smoothing
 method to a $^6$He breakup reaction at 22.5~MeV. Multistep
 breakup processes are found to be important.
}

\begin{document}
\maketitle

\section{Introduction}
\label{sec:intro}

Recent developments in radioactive beam experiments have made it
possible to study unstable nuclei away from the stability line.
Such nuclei have exotic properties, e.g., the halo
structure~\cite{Tanihata1,Tanihata2,Hansen}, in which weakly bound
valence neutrons extend far from a core nucleus.
Borromean nuclei such as $^6$He and $^{11}$Li are typical examples of
halo nuclei and are described well by a three-body
($n$+$n$+core) model. In the scattering of a three-body projectile,
it easily breaks up into its constituents, and hence the
reaction should be described as four-body ($n$+$n$+core+target)
scattering.

Thus far, many experiments have been conducted on the scattering of $^6$He from
heavy~\cite{Aumann,Wang-MSU,Chulkov,Kakuee,exp6He1,exp6He2,Sanchez} to
light targets~\cite{Aumann,Wang-MSU,Lapoux,Milin}
at high~\cite{Aumann,Wang-MSU,Chulkov,Lapoux} and
low incident energies~\cite{Milin,Kakuee,exp6He1,exp6He2,Sanchez}. \
From the theoretical point of view, it is quite difficult to solve
four-body scattering exactly.
At higher incident energies, therefore, the scattering has been
analyzed using approximate methods such as the Glauber
model~\cite{Al-Khalili1,Al-Khalili2}, adiabatic
approximation~\cite{Johnson, Christley}, multiple
scattering expansion~\cite{Crespo} and the four-body
DWBA~\cite{Chatterjee,Ershov2,Ershov4}. \
These methods are, however, not applicable to scattering at
low incident energies.

One of the most useful and reliable methods for the scattering in a wide
range of incident energies is the method of continuum-discretized coupled
channels (CDCC)~\cite{CDCC-review1,CDCC-review2}. \ In CDCC, the total
scattering wave function is expanded in terms of the complete set of
bound and continuum states of the projectile. The continuum states are
classified by linear and angular momenta, $k$ and $\ell$,
respectively; each of them is truncated at a certain value.
The $k$-continuum is then divided into small bins and the
continuum states in each bin are averaged into a single state. This
procedure of discretization is called the average method.
The $S$-matrix elements calculated with CDCC converge as the model space
is extended~\cite{CDCC-convergence-1,CDCC-convergence-2}. \
The converged CDCC solution is the unperturbed solution of
the distorted Faddeev equations, and corrections to the solution are
negligible within the region of space in which the reaction takes
place~\cite{CDCC-foundation1,CDCC-foundation2}.

Conventional CDCC based on the average method has been applied only
to three-body breakup processes in the scattering of two-body
projectiles. It is thus called {\it three-body CDCC}. For four-body
breakup processes in the scattering of three-body projectiles, the average
method is not feasible, since it requires exact three-body continuum
states of the projectile that are quite difficult to obtain.
This problem can be
circumvented by the pseudostate discretization method instead of
the average method.
In the pseudostate
discretization method, the continuum states are replaced with pseudostates
obtained by diagonalizing the internal Hamiltonian of the projectile in
a space spanned by $L^2$-type basis functions. One can adopt the
Gaussian~\cite{Matsumoto,Egami} or transformed harmonic oscillator
(THO)~\cite{Moro1} basis as the $L^2$-type basis functions.
The
validity of the pseudostate discretization method was confirmed in
the scattering of two-body projectile by the good agreement
between the CDCC solutions obtained by the pseudostate discretization and
the average methods.

The pseudostate discretization method makes CDCC applicable to the
scattering of three-body projectiles.
In fact, {\it four-body CDCC} based on the pseudostate discretization
method with Gaussian~\cite{Matsumoto2, Matsumoto3,Matsumoto4} or
THO~\cite{THO-CDCC} basis functions has been successful in describing
the elastic scattering of a three-body projectile at not only high
energies but also low energies near the Coulomb barrier.
This shows that the back-coupling effects of four-body breakup channels
on the elastic scattering, i.e., {\it virtual} four-body breakup
processes, are described well by four-body CDCC.
Further development is, however, necessary to apply four-body CDCC
to {\it real} four-body breakup processes, as shown below.

\begin{figure}[tb]
\begin{center}
 \includegraphics[clip,width=0.6\textwidth]{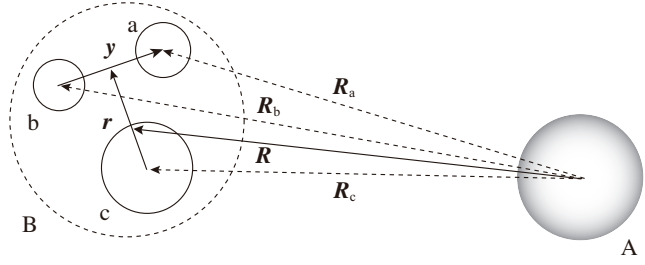}
 \caption{Illustration of the four-body system;
 A is the target and B is the projectile composed of
 the constituents a, b, and c. }
 \label{fig:4B-CDCC}
\end{center}
\end{figure}

Let us consider the four-body reaction system shown in
Fig.~\ref{fig:4B-CDCC}. The three-body projectile B, which consists
of a, b, and c, is incident on the target nucleus A.
In the final stage of the breakup reaction considered,
four particles, a, b, c, and A, are emitted.
The cross section of the reaction is described by the quintuple
differential cross section
\begin{align}
 \frac{d^5\sigma}{d\ve_a d\ve_b  d\Omega_a d\Omega_b d\Omega_c}
 \propto |S(\bm{k},\bm{K},\bm{P})|^2,
 \label{eq:quintet-x}
\end{align}
where $\ve_a$ ($\ve_b$) is the energy of the emitted particle a (b),
and $\Omega_a$, $\Omega_b$, and $\Omega_c$ are
the scattering angles of a, b, and c, respectively.
This cross section is calculated using the {\it true}
breakup $S$-matrix elements
$S(\bm{k},\bm{K},\bm{P})$ for the transition of B from the
ground states to the continuum states classified with three momenta;
$\bm{k}, \bm{K}$, and $\bm{P}$ are, respectively, conjugate to the
coordinates $\bm{r}, \bm{y}$, and $\bm{R}$ shown
in Fig.~\ref{fig:4B-CDCC}.
The single differential cross
section $d\sigma/d\ve$ with respect to the intrinsic energy $\ve$ of B
is obtained by integrating Eq.~\eqref{eq:quintet-x} over the five variables
$\ve_a$, $\ve_b$, $\Omega_a$, $\Omega_b$, and $\Omega_c$ with
energy and momentum conservations. However, the
breakup $S$-matrix
elements $\hat{S}$ calculated using CDCC are discrete functions of $\ve$,
because of pseudostate discretization, i.e.,
the diagonalization of the internal Hamiltonian of the projectile.
Thus, we need a smoothing method for deriving continuous breakup $S$-matrix
elements $S(\bm{k},\bm{K},\bm{P})$ from $\hat{S}$.

In this paper, we propose a new smoothing method for constructing
$S(\bm{k},\bm{K},\bm{P})$ from $\hat{S}$ to complete four-body CDCC.
First, we test the smoothing method in breakup
reactions of a two-body projectile and
confirm that the breakup $S$-matrix elements calculated
by the smoothing method agree with the {\lq\lq}exact'' ones,
i.e., the results with the average method of discretization.
Next, we apply the smoothing method to
the electric dipole (E1) transition of $^6$He
to its continuum states and confirm that the calculated E1
transition strength, with an additional smearing procedure
concerning the experimental resolution,
converges as the model space is extended.
Finally, we apply the smoothing method to four-body breakup reactions
of $^6$He by $^{209}$Bi at 22.5~MeV and show that
multistep breakup processes are significant.

This paper is constructed as follows. In \S\ref{sec:formulation}, we
recapitulate four-body CDCC and present a new smoothing method.
Test calculations of the smoothing method are shown in
\S\ref{sec:test} and its application to
the $^6$He+$^{209}$Bi scattering at 22.5~MeV is shown in \S\ref{sec:bux}.
Section~\ref{sec:summary} gives the summary.

\section{Formulation}
\label{sec:formulation}

\subsection{Four-body CDCC}
\label{sub:4bcdcc}

We start with the four-body Hamiltonian of
the four-body system shown in Fig.~\ref{fig:4B-CDCC}:
\begin{align}
 H &=T_R+U+H_{\rm B},\\
 U &=U_a(\bm{R}_a)+V_a^{\rm Coul}(\bm{R}_a)
    +U_b(\bm{R}_b)+V_b^{\rm Coul}(\bm{R}_b)
    +U_c(\bm{R}_c)+V_c^{\rm Coul}(\bm{R}_c),\\
 H_{\rm B}&=T_r+T_y+V_{bc}+V_{ca}+V_{ab},
\end{align}
where $\bm{R}$ is the relative coordinate between the center of masses
(cm) of B and A; we assume A to be a structureless and inert nucleus.
Note that back-coupling effects of inelastic channels concerning the
excitation of A are taken into account with the use of the optical potential
between A and each constituent of the projectile, as will be mentioned later.
The coordinate $\bm{R}_{\rm x}$ (x $=$ a, b, and c) denotes the relative
coordinate between x and A, and $T_{\xi}$ ($\xi=\bm{R}$, $\bm{r}$, and
$\bm{y}$) is the kinetic-energy operator associated with $\xi$.
The interaction $V_{\rm xx'}$ is the potential
between x and x$'$, and
$U_{\rm x}$ is the optical potential
between x and A; the Coulomb part $V_{\rm x}^{\rm Coul}$ is also
accurately treated in order to describe Coulomb breakup processes.

The basic assumption of four-body CDCC is that
the four-body reaction takes place in the model
space~\cite{Matsumoto3,Matsumoto4}
\begin{align}
 {\cal P}&=\sum_{i}|\hat{\Phi}_{i}\rangle \langle\hat{\Phi}_{i}|,
 \label{eq:com-set}
\end{align}
where $i$ denotes a set of quantum numbers, i.e.,
the energy index of the pseudostates
$n$, the total spin of the projectile $I$, and its projection on the
$z$-axis $m$. The pseudostate $\hat{\Phi}_{i}
(\equiv \hat{\Phi}_{nIm})$ satisfies
\begin{align}
\langle \hat{\Phi}_{i}|H_{\rm B}| \hat{\Phi}_{i'} \rangle =
\delta_{ii'}\hat{\ve}_{i},
\end{align}
where $\hat{\ve}_{i}$ is the eigenenergy of $\hat{\Phi}_{i}.$
The validity of this assumption is justified by the fact that
the calculated elastic and total breakup
cross sections of the four-body scattering
converge as the model space is
extended~\cite{Matsumoto, Egami, Matsumoto2,Matsumoto3, Matsumoto4}. \
One may thus regard $\{\hat{\Phi}_i\}$ as a complete set
in describing the reaction process considered. We henceforth
call $\{\hat{\Phi}_i\}$ the approximate complete set in this
meaning.

Following the discussion above, four-body CDCC starts with the four-body
Schr\"odinger equation in the model space ${\cal P}$:
\begin{align}
{\cal P}[H-E_{\rm tot}]{\cal P}|\Psi^{\rm CDCC} \rangle =0,
\label{eq:4b-Schr}
\end{align}
where $E_{\rm tot}$ is the total energy of the system.
The four-body wave function $\Psi^{\rm CDCC}$
is expanded by the approximate
complete set $\{\hat{\Phi}_{i}\}$:
\begin{align}
 |\Psi^{\rm CDCC} \rangle
 = \sum_{i} | \hat{\Phi}_{i}, \hat{\chi}_{i}\rangle,
 \label{eq:CDCC-1}
\end{align}
where $|\hat{\Phi}_{i},\hat{\chi}_{i}\rangle
=|\hat{\Phi}_{i}\rangle\otimes|\hat{\chi}_{i}\rangle$,
and $i=0$ denotes the elastic channel and others ($i\ne 0$)
the breakup channels.
The expansion coefficient $|\hat{\chi}_{i}\rangle$ describes
the relative motion between B (in state $\hat{\Phi}_{i}$)
and A.
The intrinsic energy $\hat{\ve}_{i}$ of B
and the relative momentum $\hat{P}_{i}$ between B and A
satisfy the energy conservation
\begin{align}
 E_{\rm tot}=E_{\rm in}^{\rm cm}+\hat{\ve}_{0}
=\frac{\hbar^2}{2\mu}\hat{P}_{i}^2+\hat{\ve}_{i},
\label{eq:energy-4B}
\end{align}
where $\mu$ is the reduced mass between B and A, and
$E_{\rm in}^{\rm cm}$ is the incident energy of B
in the cm system,
i.e., $E_{\rm in}^{\rm cm}\equiv{\hbar^2\hat{P}_{0}^2}/(2\mu)$.

Multiplying Eq.~\eqref{eq:4b-Schr} by
$\langle\hat{\Phi}_{i}|$ from the left leads to a set of coupled
differential equations for $|\hat{\chi}_{i}\rangle$, called the CDCC
equation,
\begin{align}
 [T_R+\hat{U}_{i,i}-E_i]|\hat{\chi}_{i}\rangle
 = -\sum_{i'\ne i} \hat{U}_{i,i'}|\hat{\chi}_{i'}\rangle,
 \label{eq:CDCC-eq}
\end{align}
where $E_i \equiv \hbar^2\hat{P}^2_i/(2\mu)
 =E_{\rm tot}-\hat{\ve}_{i}$.
The coupling potential $\hat{U}_{i,i'}$ is defined by
\begin{align}
 \hat{U}_{i,i'}
 &=\langle \hat{\Phi}_{i} | U | \hat{\Phi}_{i'}\rangle.
 \label{eq:4B-CP}
\end{align}
The CDCC equation \eqref{eq:CDCC-eq} is solved under the usual
boundary condition for $\langle \bm{R}|\hat{\chi}_i\rangle\equiv\hat{\chi}_i(\bm{R})$
\cite{CDCC-review1,CDCC-review2}.

\subsection{Pseudostate Discretization Method}
\label{sub:Pseudostates}

\begin{figure}[tb]
 \begin{center}
  \includegraphics[clip,width=0.70\textwidth]{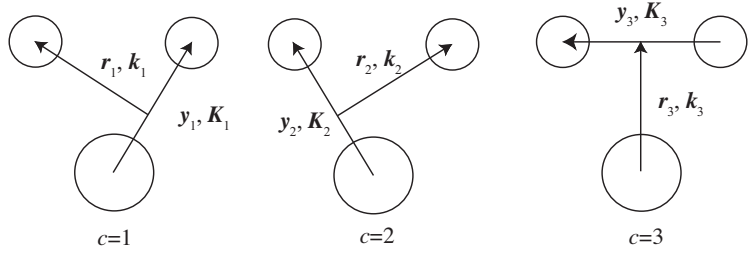}
  \caption{Jacobi coordinates of the three rearrangement channels
  ($c=$1\mbox{--}3) in the three-body system.
  Vectors ($\bm{k}_c, \bm{K}_c$) denote linear momenta
  conjugate to the coordinates ($\bm{r}_c, \bm{y}_c$).}
  \label{fig:Jacobi}
 \end{center}
\end{figure}%
In the pseudostate discretization
method~\cite{Matsumoto, Egami, Matsumoto2,Matsumoto3, Matsumoto4},
the pseudostates \{$\Hat{\Phi}_{i}$\} are
obtained by diagonalizing the internal Hamiltonian $H_{\rm B}$ of B
in a space spanned by $L^2$-type Gaussian basis
functions~\cite{H-Ka-Ki}. \
The eigenstates of $H_{\rm B}$ with negative and positive energies
correspond to the bound state(s) and pseudostates, respectively,
and the latter are regarded as discretized continuum states as mentioned
above.
Now we consider $^6$He as a projectile.
The eigenstate $\hat{\Phi}_{i}\equiv \hat{\Phi}_{nIm}$ is then written as
\begin{equation}
\langle \bm{r},\bm{y}|\hat{\Phi}_{nIm}\rangle
={\sum_{c=1}^3} \psi^{(c)}_{nIm}(\bm{r}_c,\bm{y}_c),
\end{equation}
where $c$ denotes a set of Jacobi coordinates defined in
Fig.~\ref{fig:Jacobi}.
Each $\psi^{(c)}_{nIm}$ is a sum of the Gaussian basis functions:
\begin{align}
 \psi^{(c)}_{nIm}(\bm{r}_c,\bm{y}_c)
 &= \sum_{\alpha} A_{\alpha}^{(c)nI}
 \varphi^{(c)}_{\alpha}(\bm{r}_c,\bm{y}_c), \\
 \varphi^{(c)}_{\alpha}(\bm{r}_c,\bm{y}_c)
 &= \mathcal{N}_{i\ell}\mathcal{N}_{j\lambda}
 r_c^{\ell} e^{-\nu_ir_c^2}
 y_c^{\lambda} e^{-\lambda_jy_c^2}
 \notag\\
 &\quad \times
 \Big[ \big[Y_{\ell}(\Omega_{r_c})  \otimes
 Y_{\lambda}(\Omega_{y_c})  \big]_\Lambda
 \otimes
 \big[\eta_{{1}/{2}}^{(n_1)}\otimes\eta_{{1}/{2}}^{(n_2)}\big]_{S}
 \Big]_{Im},
 \label{gauex}
\end{align}
where $\alpha$ is the abbreviation of $\{i,j,\ell,\lambda,\Lambda, S\}$;
$\ell$ ($\lambda$) is the orbital angular momentum associated with the
coordinate $\bm{r}_c$ ($\bm{y}_c$)
and $\eta_{1/2}$ is the spin wave function of each valence neutron,
$n_1$ or $n_2$. The normalization coefficients
$\mathcal{N}_{i\ell}$ and $\mathcal{N}_{j\lambda}$
are determined so as to satisfy
$ \langle \mathcal{N}_{i\ell}r_c^{\ell}\, e^{-\nu_ir_c^2}
|  \mathcal{N}_{i\ell}r_c^{\ell}\, e^{-\nu_ir_c^2}\rangle=1$
and
$ \langle  \mathcal{N}_{j\lambda}y_c^{\lambda}\,
e^{-\lambda_jy_c^2} |  \mathcal{N}_{j\lambda}y_c^{\lambda}\,
e^{-\lambda_jy_c^2} \rangle=1$.
The quantum numbers
$\ell$, $\lambda$, and $\Lambda$ are truncated by the upper limit values
$\ell_{\rm max}$, $\lambda_{\rm max}$, and $\Lambda_{\max}$, respectively.
The total spin $S$ is either 0 or 1.
The Gaussian range parameters are given
geometric progression:
\begin{align}
 \nu_i=1/{r}_{i}^2, \quad
 {r}_{i}&={r}_1
 \left({{r}_{\max}}/{{r}_{1}}\right)^{(i-1)/(i_{\rm max}-1)},
\label{Grange1}
 \\
\lambda_j=1/{y}_{j}^2, \quad
 {y}_{i}&={y}_1
 \left({{y}_{\max}}/{{y}_{1}}\right)^{(j-1)/(j_{\rm max}-1)}.
\end{align}
The states $\hat{\Phi}_{nIm}$ are
antisymmetric under the interchange between $n_1$ and $n_2$, and
hence they must satisfy
$A_{\alpha}^{({2})nI} =(-)^SA_{\alpha}^{({1})nI}$
and for $c=3$ $(-)^{\lambda+S}=1 $.
Meanwhile, the exchange between each valence neutron and each
nucleon in $^4$He is treated approximately by the orthogonality
condition model~\cite{OCM}. \ The eigenenergies
$\hat{\ve}_{nI}$ of $^6$He
and the corresponding expansion coefficients
$A_{\alpha}^{(c)nI}$ are
determined by diagonalizing $H_{\rm B}$~\cite{GEM6He1,GEM6He}. \
In the diagonalization procedure
one needs the coordinate transformation between
the rearrangement channels. Details of this transformation
are described in Appendix \ref{A1}.

\subsection{Smoothing Method}
\label{sub:smoothing}

The exact breakup $T$-matrix elements to the continuum state
$\psi(\bm{k},\bm{K})$ of B are given by
\begin{align}
 T^{\rm EX}(\bm{k}, \bm{K},\bm{P})
 &=\langle \psi(\bm{k},\bm{K}), P|U|\Psi\rangle, \
 \label{eq:t-mat}
\end{align}
with $ |\psi(\bm{k},\bm{K}), P\rangle =
|\psi(\bm{k},\bm{K})\rangle\otimes|P\rangle$, where $\bm{k}$, $\bm{K}$,
and $\bm{P}$ are the momenta in the asymptotic region associated
with the coordinates $\bm{r}$, $\bm{y}$, and $\bm{R}$, respectively;
$|\psi(\bm{k},\bm{K})\rangle$ is the exact three-body wave function of
B with the energy $\ve$ satisfying
\begin{align}
[H_{\rm B}-\ve]|\psi(\bm{k},\bm{K})\rangle=0,
\label{eq:Sch-B}
\end{align}
and $|P\rangle$ is the plane wave function satisfying
\begin{align}
  [T_R-(E_{\rm tot}-\ve)]|P\rangle =0.
\end{align}
The exact four-body (a+b+c+A) wave function $\Psi$
can be replaced with the corresponding CDCC wave function
$\Psi^{\rm CDCC}$ with good accuracy.
Inserting the approximate complete set
${\cal P}$ of Eq.~\eqref{eq:com-set}
between the bra vector and the operator $U$ on the
right-hand side of Eq.~\eqref{eq:t-mat}, we can obtain the
approximate smooth $T$-matrix elements $T(\bm{k}, \bm{K}, \bm{P})$:
\begin{align}
 T(\bm{k},\bm{K}, \bm{P})
 &=
 \sum_{i}\langle \psi(\bm{k},\bm{K})|\hat{\Phi}_{i}\rangle
 \langle\hat{\Phi}_{i}, \hat{P}_i | U |\Psi^{\rm CDCC} \rangle\notag\\
 &\equiv \sum_{i}\mathcal{F}_{i}(\bm{k},\bm{K})  \hat{T}_{i},
\label{eq:t-mat2}
\end{align}
where
$|\hat{\Phi}_{i}, \hat{P}_i \rangle=
|\hat{\Phi}_{i}\rangle\otimes|\hat{P}_i\rangle$,
$\mathcal{F}_{i}(\bm{k},\bm{K})$ is the smoothing factor defined by
\begin{align}
  \mathcal{F}_{i}(\bm{k},\bm{K})
 &=\langle \psi(\bm{k},\bm{K})|\hat{\Phi}_{i}\rangle,
\label{eq:smoothing}
\end{align}
and $\hat{T}_{i}$ is the breakup $T$-matrix element of CDCC
defined by
\begin{align}
 \hat{T}_{i}
 &=\langle\hat{\Phi}_{i}, \hat{P}_i | U | \Psi^{\rm CDCC}\rangle.
\end{align}
Since the breakup $T$-matrix elements are
proportional to the breakup $S$-matrix elements,
Eq.~\eqref{eq:t-mat2} is reduced to
\begin{align}
 S(\bm{k},\bm{K},\bm{P})
 = \sum_{i}\mathcal{F}_{i}(\bm{k},\bm{K}) \hat{S}_{i}.
 \label{eq:smooth-S}
\end{align}

The smoothing factor $\mathcal{F}_{i}(\bm{k},\bm{K})$
is the overlap between the pseudostate $\hat{\Phi}_{i}$ and
the exact continuum state $\psi(\bm{k},\bm{K})$.
In principle, the exact continuum
state is obtained by solving the three-body Schr\"odinger
equation \eqref{eq:Sch-B} under the initial condition that the three
particles a, b, and c, are incident particles. In practice, however, it
is quite difficult to do so.
Therefore, we approximately solve the equation in the model space ${\cal P}$
of Eq.~\eqref{eq:com-set}:
\begin{align}
 {\cal P}|\psi\rangle&={\cal P}|\psi_0\rangle
 +{\cal P}\frac{1}{\ve-T_{r}-T_{y}+i\epsilon}{\cal P}V{\cal P}
 |\psi\rangle,
 \label{eq:LS-2}
\end{align}
where $V=V_{ab}+V_{bc}+V_{ca}$ and $\psi_0$ describes free propagation
of the three incident particles satisfying
\begin{align}
 \left[T_r+T_y-\ve\right]|\psi_0\rangle&=0.
\end{align}
The solution to Eq.~\eqref{eq:LS-2} should converge as the model space is
extended.
This is tested in \S\ref{sec:test}.
Multiplying Eq.~\eqref{eq:LS-2} by $\langle\hat{\Phi}_i|$ leads to
\begin{align}
\mathcal{F}_{i}(\bm{k},\bm{K})
 =\tilde{\Phi}_{i}(\bm{k},\bm{K}) +\sum_{jk} G_{ij} V_{jk}
 \mathcal{F}_{k}(\bm{k},\bm{K}),
 \label{eq:sf}
\end{align}
where
\begin{align}
 \tilde{\Phi}_{i}(\bm{k},\bm{K})
 &= \langle \hat{\Phi}_i | \psi_0\rangle
 = \langle \psi_0 | \hat{\Phi}_i \rangle,
 \label{eq:mat-p-l2}\\
 G_{ij}
 &=\langle\hat{\Phi}_i|\frac{1}{\ve-T_{r}-T_{y}+i\epsilon}
 |\hat{\Phi}_j\rangle,
 \label{eq:mat-gr2}\\
 V_{jk}
 &=\langle \hat{\Phi}_j|V|\hat{\Phi}_k\rangle.
\end{align}
Since Eq.~\eqref{eq:sf} is a set of linear equations
for $\mathcal{F}_{i}(\bm{k},\bm{K})$, one can easily obtain
a solution for $\mathcal{F}_{i}(\bm{k},\bm{K})$ once
$\tilde{\Phi}_{i}(\bm{k},\bm{K})$, $G_{ij}$, and
$V_{jk}$ are evaluated. We emphasize here that what we need
in the present smoothing method is not
the exact three-body continuum wave function $\psi(\bm{k},\bm{K})$
itself but the smoothing factors $\mathcal{F}_{i}(\bm{k},\bm{K})$.
In other words, as shown in Eq.~\eqref{eq:smoothing},
we need only the overlap between $\psi(\bm{k},\bm{K})$ and
$\hat{\Phi}_{i}$; the latter is an accurate three-body wave
function in the model space that is significant for the
four-body reaction under consideration.
Thus, henceforth, we call this procedure based on model space
truncation
{\it the model space smoothing method}.
The six-fold integral over $\bm{r}$ and $\bm{y}$ in Eq.~\eqref{eq:mat-p-l2}
can be made analytically, and the six-fold integral
in Eq.~\eqref{eq:mat-gr2} is
reduced to a single integral.
These properties markedly simplify numerical calculations.
We show in Appendix \ref{A2} the explicit form of
$\tilde{\Phi}_{i}(\bm{k},\bm{K})$, i.e., the Fourier transform
of $\hat{\Phi}_i$, and in Appendix \ref{A3} that of $G_{ij}$.

We remark that in $\mathcal{F}_{i}(\bm{k},\bm{K})$,
or equivalently, in Eq.~\eqref{eq:LS-2}, the rearrangement channels of
the three particles are fully taken into account. On the
other hand, in the calculation of scattering processes
by four-body CDCC, we neglect four-body rearrangement channels,
in which at least one constituent of the projectile is bound in
the target nucleus. The contribution of the rearrangement
channels is, however, theoretically shown to be negligible in
forward angle scattering.~\cite{CDCC-foundation1,CDCC-foundation2} \
Recently, it has also been confirmed numerically by comparing
the results of three-body CDCC for deuteron elastic and breakup
processes by $^{12}$C at 56 MeV with the exact solution of
the Faddeev equation.~\cite{Fonseca}
Such a comparison for four-body scattering processes will be
very interesting.

\section{Test Calculations}
\label{sec:test}

\subsection{Breakup Reaction of Two-body Projectile}
\label{sub:test1}

\begin{figure}[tb]
 \begin{center}
  \includegraphics[clip,width=0.7\textwidth]{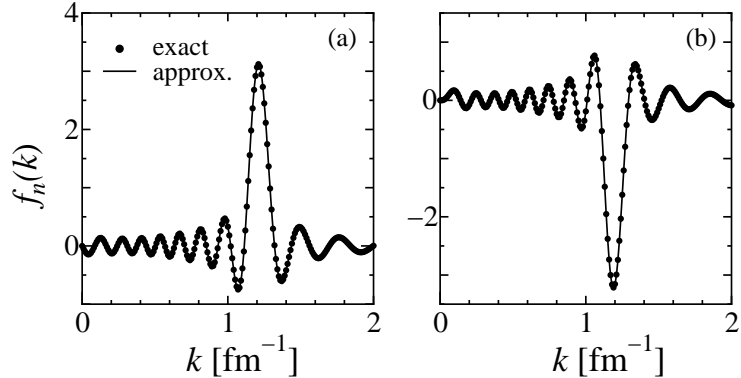}
  \caption
  {Smoothing factor as a function of
  $k$ for $n=16$. The left (right) panel
  corresponds to $\ell=0$ ($\ell=2$).
  The dots and solid curve denote the results of
  the exact calculation
  and those obtained by the model space smoothing method, respectively.}
  \label{fig:s-fac2}
 \end{center}
\end{figure}%

\begin{figure}[tb]
 \begin{center}
  \includegraphics[clip,width=0.7\textwidth]{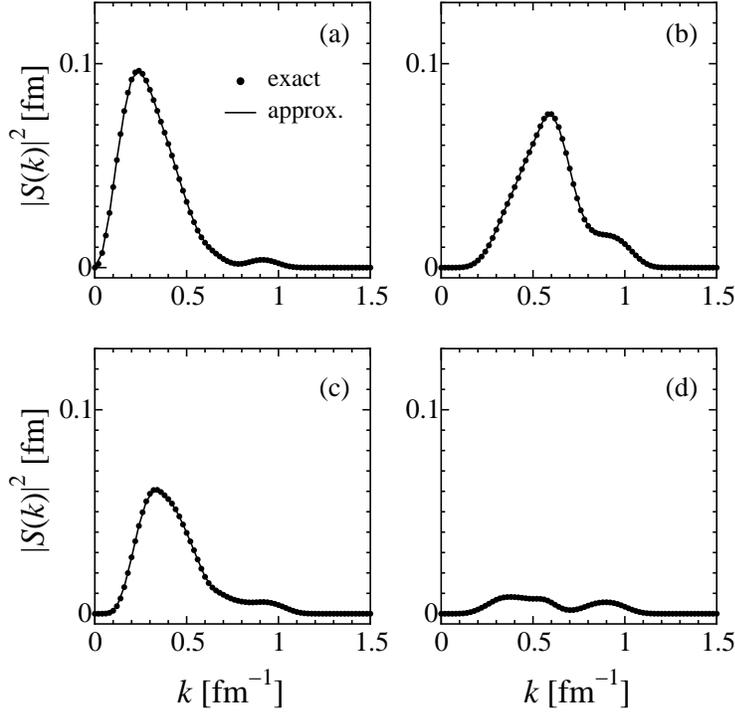}
  \caption{Squared magnitudes of the smoothed breakup
  $S$-matrix elements
  $|S(k)|^2$ as a function of $k$ at the total angular momentum $J=17$ for
  the $d$+$^{58}$Ni scattering at 80~MeV.
  Panels (a), (b), (c), and (d) correspond to $(\ell, L)=(0, 17)$,
  $(2, 15)$, $(2, 17)$, and $(2, 19)$, respectively.
  The results of the exact calculation (model space smoothing method)
  are shown by the dots (solid curve). }
  \label{fig:S2-d2}
 \end{center}
\end{figure}%

In this subsection, we consider the $^{58}$Ni$(d, pn)$ reaction at 80~MeV to
test the validity of the model space smoothing method. In this reaction,
the projectile ($d$) consists of a proton ($p$) and a neutron ($n$), and the
total system is composed of $p$, $n$, and A. The Coulomb
breakup processes are neglected, i.e., Coulomb potential is treated
as a function of the relative coordinate $\bm{R}$ between
the cm of $d$ and $^{58}$Ni.
We adopt the model space smoothing method described in
\S\ref{sub:smoothing} to obtain continuous breakup $S$-matrix elements
for this reaction. Note that, in this case,
the coordinate $\bm{y}$ and its conjugate momentum
$\bm{K}$ do not appear and $\bm{r}$ denotes the relative coordinate
between $p$ and $n$.
As for the interaction between $p$ and $n$,
we cite that in Ref.~\citen{Matsumoto}. The ground and breakup states of
the $p$+$n$ subsystem, classified with the linear and angular momenta $k$
and $\ell$, respectively, are obtained by diagonalizing the Hamiltonian
of a subsystem with complex-range Gaussian basis
functions~\cite{H-Ka-Ki}. As for the parameter set of the basis functions,
we take \{$2N_{\max}=40$, $r_{1}=1.0$~fm, $r_{\max}=20$~fm\},
where $2N_{\max}$ is the total number of the
basis functions, and $r_{1}$ and $r_{\max}$ are the Gaussian range
parameters given in Eq.~\eqref{Grange1}.
The maximum momentum $k_{\max}$
is taken to be 1.3 fm$^{-1}$, and the number of states is 17 for
each of $\ell=0$ and $2$. The optical potentials of $n$-$^{58}$Ni and
$p$-$^{58}$Ni systems are cited from Ref.~\citen{Matsumoto}. \
In solving the CDCC equation, the scattering wave functions
are connected with the asymptotic forms at $R_{\max}=30$~fm.

When a two-body projectile such as $d$ is considered,
it is possible to obtain an exact smoothing factor
without making the model space truncation in Eq.~\eqref{eq:LS-2}.
In Fig.~\ref{fig:s-fac2},
the approximate smoothing factor with
the model space smoothing method (solid curve) is compared with
the exact smoothing factor (dots) in the case of
$n=16$ with $\ell=0$ (left panel) and $2$ (right panel).
In each panel, the two results agree very well with each other.
In fact, the difference between the two is below 1\% level.

Figure~\ref{fig:S2-d2} shows
the squared magnitudes of the
breakup $S$-matrix elements for
the total angular momentum $J=17$; panels (a), (b), (c), and (d)
correspond to $(\ell, L)=(0, 17)$, $(2, 15)$, $(2, 17)$,
and $(2, 19)$, respectively. Again, the results obtained by
the model space smoothing method (solid curves) agree very well
with the results of the exact calculation (dots).

\subsection{Electric Dipole Transition of $^6$He to 1$^-$ Continuum States}
\label{sub:test2}

In the calculation of the $^4$He+$n$+$n$ three-body wave
functions, we take the same Hamiltonian $H_{\rm B}$ as that in
Ref.~\citen{Matsumoto3}, which is based on the orthogonality
condition model (OCM)~\cite{OCM}. \
In the present calculation, however,
the OCM potential to exclude the forbidden state
is introduced into only the $0^+$ state.
Consequently, one can obtain the $0^+$ ground state of $^6$He
while excluding the forbidden state in the diagonalization of $H_{\rm B}$.
On the other hand, for the $1^-$ state, not only the physical
pseudostates to be regarded as discretized continuum states
of the $^4$He+$n$+$n$ three-body system, but also unphysical
eigenstates that have large overlaps with the forbidden state
are obtained, because of the absence of the OCM potential.
We exclude the latter by hand, following the variation-before-projection
method.
This rather special treatment of the $1^-$ state is
due to the following fact. In the present calculation, as one
can see below, so many Gaussian basis functions are used for
the $1^-$ state to cover a wide range of internal coordinates
concerned. In this case, when an OCM potential that has a core
of about $10^6$ MeV appears, the numerical calculation of the
diagonalization of the $^6$He internal Hamiltonian becomes
unstable.
Note that the $0^+$ pseudostates and $2^+$ state of $^6$He,
which are included in Ref.~\citen{Matsumoto3}, are not relevant to the
present calculation of the E1 strength distribution of $^6$He.

As for the basis functions taken in the diagonalization of $H_{\rm B}$,
we adopt the real-range Gaussian basis functions.
The parameter set for the $0^+$ state is shown
in Table~\ref{tab:conf-0}.
For the $1^-$ state, we take the three parameter sets
shown in Table~\ref{tab:conf-1} and see the dependence of
the E1 strength distribution on them;
the model space described by the $1^-$ pseudostates
of $^6$He taken in the calculation dictates the
convergence of the E1 distribution.
Note that for the $1^-$ states, $r_{\max}$ and $y_{\max}$ are
taken to be 50--80 fm to cover the large model space concerned.
The $S=1$ components of the $1^-$ states are found to be negligible
and not included in the present calculation.

The electric dipole (E1) strength
from the $0^+$ ground state $\Phi_{00}^{gs}$
to the $n$th $1^-$ pseudostate is
\begin{align}
 B({\rm E1}; n)
 &= N
 \sum_{\mu m}
 |\langle
 \hat{\Phi}_{n1m}|\mathcal{O}_{1\mu}({\rm E1})|\Phi_{00}^{gs}\rangle|^2,
 \label{eq:indi-BE1}
\end{align}
where $\mathcal{O}_{1\mu}({\rm E1})=r_3Y_{1\mu}(\Omega_{r_3})$ and
$N = \left({N_nN_c}/{A}\right)^2$.
The mass number and neutron number of the core nucleus
are denoted $A$ and $N_c$, respectively, and $N_n$ is
the number of valence neutrons; for $^6$He, $A=6$ and $N_n=N_c=2$.

\begin{table}[tbp]
 \begin{center}
 \caption{Parameters of the basis functions
  for the $0^{+}$ state of $^6$He.}
  \catcode `?=\active \def?{\phantom{0}}
  \begin{tabular}{cc cc cc cc cc cc cc cc cc cc c}
   \hline\hline
   \multirow{2}{*}{$c$} && \multirow{2}{*}{$\ell$} &&
   \multirow{2}{*}{$\lambda$} && \multirow{2}{*}{$\Lambda$} &&
   \multirow{2}{*}{$S$} && \multirow{2}{*}{$i_{\max}$} &&
    $r_{1}$ && $r_{\max}$ && \multirow{2}{*}{$j_{\max}$} &&
   $y_{1}$ && $y_{\max}$  \\
    &&  &&  &&  &&  &&  &&(fm) && (fm) &&  && (fm) && (fm)  \\
   \hline
   3  && 0 && 0 && 0 && 0 && 10 && 0.1 && 20.0 && 10 && 0.5 && 20.0 \\
   1,2  && 0 && 0 && 0 && 0 && 10 && 0.5 && 20.0 && 10 && 0.5 && 20.0 \\
   3  && 1 && 1 && 1 && 1 && 10 && 0.1 && 20.0 && 10 && 0.5 && 20.0 \\
   1,2  && 1 && 1 && 0 && 0 && 10 && 0.5 && 20.0 && 10 && 0.5 && 20.0 \\
   1,2  && 1 && 1 && 1 && 1 && 10 && 0.5 && 20.0 && 10 && 0.5 && 20.0 \\
   \hline
  \end{tabular}
  \label{tab:conf-0}
 \end{center}
\end{table}%
\begin{table}[tbp]
\begin{center}
 \caption{Three parameter sets (I, II, and III) of the basis
 functions for the $1^-$ state of $^6$He.}
 \catcode `?=\active \def?{\phantom{0}}
 \begin{tabular}{cc cc cc cc cc cc cc cc cc cc c}
  \multicolumn{21}{l}{Set I}\\
  \hline\hline
   \multirow{2}{*}{$c$} && \multirow{2}{*}{$\ell$} &&
   \multirow{2}{*}{$\lambda$} && \multirow{2}{*}{$\Lambda$} &&
   \multirow{2}{*}{$S$} && \multirow{2}{*}{$i_{\max}$} &&
    $r_{1}$ && $r_{\max}$ && \multirow{2}{*}{$j_{\max}$} &&
    $y_{1}$ && $y_{\max}$  \\
    &&  &&  &&  &&  &&  &&(fm) && (fm) &&  && (fm) && (fm)  \\
  \hline
  3  && 0 && 1 && 1 && 0 && 17 && 0.5 && 50.0 && 17 && 0.5 && 50.0 \\
  1,2  && 0 && 1 && 1 && 0 && 17 && 0.5 && 50.0 && 17 && 0.5 && 50.0 \\
  1,2  && 1 && 0 && 1 && 0 && 17 && 0.5 && 50.0 && 17 && 0.5 && 50.0 \\
  \hline\hline
  \multicolumn{21}{l}{Set II}\\
  \hline
  3  && 0 && 1 && 1 && 0 && 18 && 0.5 && 70.0 && 18 && 0.5 && 70.0 \\
  1,2  && 0 && 1 && 1 && 0 && 18 && 0.5 && 70.0 && 18 && 0.5 && 70.0 \\
  1,2  && 1 && 0 && 1 && 0 && 18 && 0.5 && 70.0 && 18 && 0.5 && 70.0 \\
  \hline\hline
  \multicolumn{21}{l}{Set III}\\
  \hline
  3  && 0 && 1 && 1 && 0 && 19 && 0.5 && 80.0 && 19 && 0.5 && 80.0 \\
  1,2  && 0 && 1 && 1 && 0 && 19 && 0.5 && 80.0 && 19 && 0.5 && 80.0 \\
  1,2  && 1 && 0 && 1 && 0 && 19 && 0.5 && 80.0 && 19 && 0.5 && 80.0 \\
  \hline
 \end{tabular}
 \label{tab:conf-1}
\end{center}
\end{table}%

\begin{figure}[tb]
 \begin{center}
  \includegraphics[clip,width=0.50\textwidth]{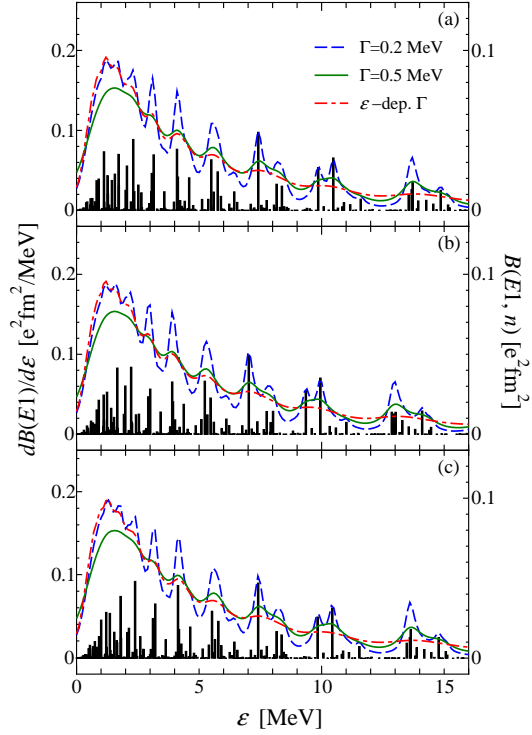}
  \caption{E1 strength distribution of $^6$He
  as a function of $\ve$.
  In each panel, the bars show ${B}({\rm E1};n)$ from the $0^+$ ground state
  to the pseudostates of $^6$He. Panels (a), (b), and (c)
  correspond to the results of Sets I, II, and III, respectively.
  The solid, dashed, and dot-dashed curves respectively show
  the smeared
  E1 strength distributions with $\Gamma=0.5$, 0.2, and
  $0.1(\hat{\varepsilon}_n+0.975)$.
  }
  \label{fig:BE1-1}
 \end{center}
\end{figure}%

Since the final states \{$\hat{\Phi}_{n1m}$\} of the E1 transition
have discrete energies \{$\hat{\ve}_n$\}, $B({\rm E1}; n)$
is a discrete function of the energy  $\ve$
of the $^4$He+$n$+$n$ system in the final state,
as shown in Fig.~\ref{fig:BE1-1} (the bars); panels (a), (b), and (c)
show the results with Sets I, II, and III, respectively.
$B({\rm E1}; n)$ should be smoothed in some way
to be compared with
the corresponding experimental data~\cite{Aumann, Wang-MSU}.

A possible simple way is to assume a Lorentzian function
for the $\ve$ dependence of each $B({\rm E1}; n)$~\cite{hagino}:
\begin{align}
 \frac{dB({\rm E1})}{d\ve}
&=\sum_{n}\frac{\Gamma}{\pi}\frac{1}{(\ve-\hat{\ve}_n)^2+\Gamma^2}B({\rm E1};n).
\label{eq:simple-smoothing}
\end{align}
In Fig.~\ref{fig:BE1-1}, the solid (dashed) curves show the results of
Eq.~\eqref{eq:simple-smoothing} with a width $\Gamma$ of 0.5 ($0.2$)~MeV.
With $\Gamma=0.2$~MeV, the E1 strength has an unnatural oscillation,
which is independent of the parameter set of the basis functions.
Even if $\Gamma$ increases to 0.5~MeV, the distribution becomes smooth
only in  the small $\ve$ region of $\ve \la 3$~MeV.
Also shown by the dot-dashed curves are the results with the
energy-dependent width $\Gamma=0.1(\hat{\varepsilon}_n+0.975)$ cited from
Ref.~\citen{Aumann}. Again, an unnatural oscillation is found
in each panel.
In all the cases, unphysically, the smeared distribution is finite at $\ve=0$.
Thus, this simple way of smoothing is not helpful for obtaining a
result that can be compared with the
experimental data~\cite{Aumann, Wang-MSU}.

The exact E1 strength $B^{\rm EX}({\rm E1})$ is defined by
\begin{align}
  B^{\rm EX}({\rm E1}) =  N  \iint d\bm{k}d\bm{K} \sum_{\mu m}
 |\langle \psi_{1m}(\bm{k}, \bm{K})|
 \mathcal{O}_{1\mu}({\rm E1})|\Phi_{00}^{gs}\rangle|^2,
 \label{eq:BE1-1}
\end{align}
where $\psi_{1m}(\bm{k}, \bm{K})$ is the exact continuum state of $1^-$
with the momenta $\bm{k}$ and $\bm{K}$. The E1 strength distribution function
${dB^{\rm EX}({\rm E1})}/{d\ve}$ is obtained using
\begin{align}
 \frac{dB^{\rm EX}({\rm E1})}{d \ve}
 &= N
 \iint d\bm{k}d\bm{K}
 \delta\left(\ve-\frac{\hbar^2k^2}{2\mu_r}-\frac{\hbar^2K^2}{2\mu_y}
 \right)
\sum_{\mu m}
 |\langle \psi_{1m}(\bm{k},\bm{K})|\mathcal{O}_{1\mu}({\rm E1})
 |\Phi_{00}^{gs}\rangle|^2.
 \label{eq:BE1-dist2}
\end{align}
Inserting the approximate complete set ${\cal P}$ of Eq.~\eqref{eq:com-set}
between the
bra vector $\langle \psi_{1m}(\bm{k},\bm{K}) |$ and the operator
$\mathcal{O}_{1\mu}({\rm E1})$ on the right hand side of Eq.~\eqref{eq:BE1-dist2} leads to
the E1 strength distribution  ${dB({\rm E1})}/{d\ve}$
based on the model space smoothing method:
\begin{align}
 \frac{dB({\rm E1})}{d\ve}
 & = N
  \iint d\bm{k}d\bm{K}
 \delta
 \left(\ve-\frac{\hbar^2k^2}{2\mu_r}-\frac{\hbar^2K^2}{2\mu_y}\right)
 \sum_{\mu m}
 \Big|\sum_{n}\mathcal{F}_{n1m}(\bm{k},\bm{K})\mathcal{M}({\rm E1}; n)
 \Big|^2, \label{eq:BE1-3}
\end{align}
where
\begin{align}
 \mathcal{M}({\rm E1}; n)
 &=
 \langle\hat{\Phi}_{n1m}|\mathcal{O}_{1\mu}({\rm E1})|\Phi_{00}^{gs}\rangle,
 \label{eq:BE1-4} \\
 \mathcal{F}_{n1m}(\bm{k},\bm{K})
 &=\langle \psi_{1m}(\bm{k},\bm{K})|\hat{\Phi}_{n1m}\rangle.
 \label{eq:BE1-5}
\end{align}
Thus, the E1 strength distribution function ${dB({\rm E1})}/{d\ve}$
is obtained from the smoothing factor
$\mathcal{F}_{n1m}(\bm{k},\bm{K})$ and the discrete E1 transition matrix
element ${\mathcal{M}({\rm E1};n)}$. Note that ${\mathcal{M}({\rm E1};n)}$
is related to the discrete E1 strength $B({\rm E1};n)$ through
Eq.~\eqref{eq:indi-BE1}. The main task to obtain
${dB({\rm E1})}/{d\ve}$ is, therefore, to evaluate the smoothing factor
$\mathcal{F}_{n1m}(\bm{k},\bm{K})$.

\begin{figure}[tb]
\begin{center}
 \includegraphics[clip,width=0.9\textwidth]{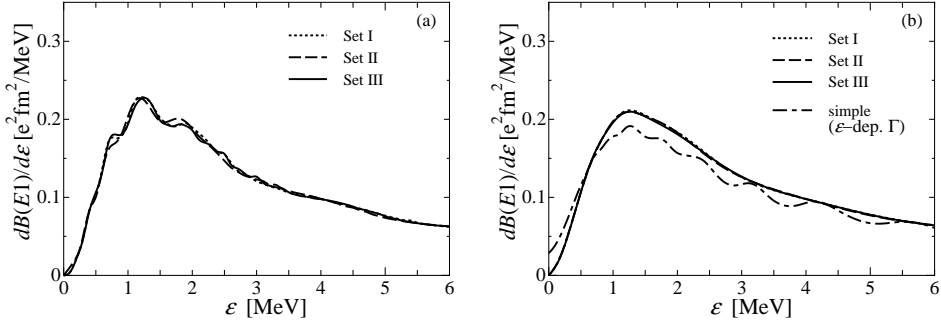}
 \caption
 {E1 strength distribution of $^6$He as a
 function of $\ve$.
 The left and right panels correspond to the nonsmeared and
 smeared results, respectively.
 In each panel, the dotted, dashed, and solid curves respectively show
 the results of Sets I, II, and III.
 Also shown for comparison by the dot-dashed curve is the result of
 the simple smoothing method with the energy-dependent $\Gamma$,
 i.e., the dot-dashed curve in Fig.~\ref{fig:BE1-1}(c).
 }
 \label{fig:BE1-2}
\end{center}
\end{figure}%

Figure~\ref{fig:BE1-2}(a) shows the dependence of ${dB({\rm E1})}/{d\ve}$
on the model space, i.e., the three parameter sets in
Table~\ref{tab:conf-1}.
The dotted, dashed, and solid curves represent the results calculated
with Sets I, II, and III, respectively.
The E1 strength distribution function almost converges
as the model space is extended. Strictly speaking, however,
the convergence is not perfect.
This is because the largest model space of the $^6$He
wave function that we can treat in the present calculation, i.e., Set III,
is still not sufficient to give a perfect convergence.
Note that we here consider the Coulomb dipole excitation of $^6$He.
Since the corresponding interaction has a long range,
the model space of $^6$He, in the $1^-$ state in particular,
requires quite large values of $r_{\max}$ and $y_{\max}$ of the
Gaussian basis functions, which considerably increases the number of
the basis functions.
On the other hand, in the nuclear breakup
process of two-body deuteron shown in \S\ref{sub:test1},
one can prepare an approximate complete set of deuteron quite easily,
because not only this breakup potential concerned is short-ranged
but also deuteron has only one internal coordinate.

Thus, at this stage the model space smoothing method cannot
give a pure theoretical result fully converged owing to
computational limits. In order to compare the theoretical
result with experimental data, however, one may use an additional smearing
procedure that takes account of experimental resolution.
This procedure is expressed by
\begin{align}
 \left\langle \frac{dB({\rm E1})}{d\ve} \right\rangle
 &=\int d\ve' w(\ve,\ve')
 \frac{dB({\rm E1})}{d\ve'}.
 \label{eq:smear}
\end{align}
As for the weight function $w(\ve,\ve')$,
we use the Lorentzian distribution with an energy-dependent width
$\Gamma=0.1(\ve'+0.975)$ cited from Ref.~\citen{Aumann}. \
Figure~\ref{fig:BE1-2}(b) represents the model space dependence of the
E1 strength distribution function smeared with Eq.~\eqref{eq:smear}.
A clear convergence is seen in the figure and the result converged
is consistent with those of the other existing
methods~\cite{Cobis,Danilin2,CSM-Myo,CSM-review}. \
This is also the case if we adopt the Gaussian distribution for
$w(\ve,\ve')$; the difference in the result from that with
the Lorentzian distribution is within the thickness of the curve
shown in Fig.~\ref{fig:BE1-2}(b), at least for $\ve \la 10$~MeV.
Also, as shown in Fig.~\ref{fig:BE1-2}(b), the dot-dashed curve
represents the result of the simple smoothing method of
Eq.~\eqref{eq:simple-smoothing} with Set III and
the energy-dependent $\Gamma$, i.e.,
the dot-dashed curve in Fig.~\ref{fig:BE1-1}(c).
One sees the dot-dashed curve deviates considerably from the solid curve.
Thus, the simple smoothing method is not accurate as mentioned above.
This is found to be also the case when the above-mentioned
$w(\ve,\ve')$ is taken in the simple smoothing method
using Eq.~\eqref{eq:simple-smoothing}.

Through the tests in \S\ref{sub:test1} and \ref{sub:test2},
one can say that the model space smoothing method is
reliable in the description of experimental data of the
breakup reactions concerned.
Therefore,
four-body CDCC based on the pseudostate discretization method and
the model space smoothing method is expected to be applicable
to four-body Coulomb and nuclear breakup reactions
such as the $(^6\textrm{He},{}^4\textrm{He}\, n\, n)$ reaction.

\section{Application to Four-Body Breakup Cross Section}
\label{sec:bux}

In this section, we apply the model space smoothing method to the breakup
reaction of $^6$He by $^{209}$Bi at 22.5~MeV.
The four-body Hamiltonian $H$ taken here is the same as that
in Ref.~\citen{Matsumoto4} except that the OCM potential is introduced
into only the $0^+$ state of $^6$He.
Moreover, since this is the first trial of four-body CDCC to
{\it real} four-body breakup
reactions, for simplicity, the model space in the present four-body CDCC
calculation is taken to be smaller than that in Ref.~\citen{Matsumoto4};
we include only the $0^+$ ground state and $1^-$ pseudostates
of $^6$He in the coupled-channel calculation.
These states are obtained as in \S\ref{sub:test2}; the
parameter set I is used for the $1^-$ state.
In CDCC calculation, the maximum internal energy $\ve_{\max}$
of $^6$He is taken to be 7~MeV, which results in 170 $1^-$ pseudostates.
The scattering waves between $^6$He and $^{209}$Bi are numerically
integrated up to $R_{\max}=200$ and connected with the asymptotic forms.
The maximum total angular momentum $J_{\max}$ is taken to be 200.

The angle-integrated breakup cross section to the continuous breakup channel
with the momenta $\bm{k}$ and $\bm{K}$ is given by
\begin{align}
 \sigma(\bm{k}, \bm{K})
 &= \sum_{J,L}\frac{\pi(2J+1)}{P_0^2}
 \Big|\sum_{Im}S_{IL}^{J}(\bm{k}, \bm{K})\Big|^2,
 \label{eq:exact-Xsec}
\end{align}
where $P_0$ is the initial momentum of $^6$He in the cm system,
$J$ is the total angular momentum
of the four-body system, and $S_{IL}^{J}(\bm{k}, \bm{K})$
is the breakup $S$-matrix element for the transition from the initial
channel to the breakup channel with $(I, L, J)$.
Here, $I$ is the total spin of the projectile, and $L$ is the orbital
angular momentum associated with $\bm{R}$.

Inserting the partial-wave expansion form of Eq.~\eqref{eq:smooth-S}
into Eq.~\eqref{eq:exact-Xsec}, one can find
\begin{align}
 \sigma(\bm{k},\bm{K})
 &=
 \sum_{J,L}\frac{\pi(2J+1)}{P_0^2}
 \Big|\sum_{nIm}
 \mathcal{F}_{nIm}(\bm{k},\bm{K})\hat{S}_{nIL}^{J}
 \Big|^2,
\end{align}
where $\hat{S}_{nIL}^{J}$ is the discrete breakup $S$-matrix
element for the transition from the initial channel to the
$n$th discrete breakup channel with the quantum numbers $(I, L, J)$.
Thus, the angle-integrated energy spectrum of the breakup reaction is
obtained by
\begin{align}
 \frac{d\sigma}{d\ve}
 =\iint d\bm{k}d\bm{K}
 \delta\left(\ve
 -\frac{\hbar^2k^2}{2\mu_r}-\frac{\hbar^2K^2}{2\mu_y}\right)
 \sigma(\bm{k},\bm{K}).
\label{eq:smoothing-x}
\end{align}
\begin{figure}[tb]
\begin{center}
 \includegraphics[clip,width=0.6\textwidth]{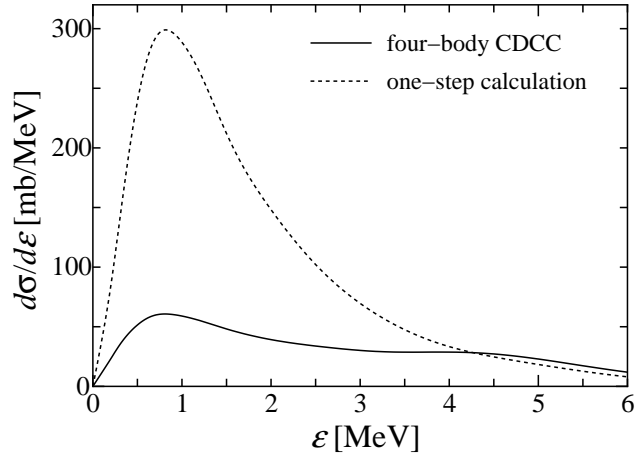}
 \caption
 { Angle-integrated energy spectrum $\langle d\sigma/d\ve\rangle$
 of $^6$He+$^{209}$Bi scattering at 22.5~MeV.
 The solid curve shows the result of the
 four-body CDCC calculation  and
 the dotted curve shows the result of the one-step calculation.
 These results have been smeared by the same procedure described in
 \S\ref{sub:test2}.
  }
 \label{fig:xsec1}
\end{center}
\end{figure}%
Since we do not obtain a fully converged theoretical result,
we calculate, as in \S\ref{sub:test2}, the smeared result with
\begin{align}
 \left\langle \frac{d\sigma}{d\ve} \right\rangle
 &=\int d\ve' w(\ve,\ve')
 \frac{d\sigma}{d\ve},\\
 \label{eq:smear2}
 w(\ve,\ve')
 &=\frac{0.1(\ve'+0.975)}{\pi}\frac{1}{(\ve-\ve')^2
   +\{0.1(\ve'+0.975)\}^2}.
\end{align}

Figure~\ref{fig:xsec1} shows the result of $\langle d\sigma/d\ve\rangle$
as a function of $\ve$.
The solid curve is the result of four-body CDCC and the dashed
curve is the result of the one-step calculation in which
only one-step transition from
the elastic channel to the individual breakup channels is
taken into account. The latter is significantly larger than the former.
This clearly shows the significant reduction in the size of the breakup
cross section due to multistep transition.
Furthermore, the two calculations give quite different
$\ve$ dependences for $\ve \ga 2$ MeV.
Therefore, the assumption of a one-step transition is not valid
at low incident energies.
The one-step assumption is often adopted when E1 strength
distribution is derived from a measured breakup cross section.
Even though this has been conducted at higher incident
energies~\cite{Aumann,Wang-MSU}, where multistep processes
are expected to be less important,
it will be very interesting and important to quantitatively
evaluate the contributions of multistep breakup processes
in this energy region.

\section{Summary}
\label{sec:summary}
We propose a new smoothing method based on model space truncation,
i.e., {\it the model space smoothing method}, for constructing
a smooth breakup cross section from a discrete one obtained CDCC.
This method allows the evaluation of not only three-body breakup
cross sections but also four-body ones, which is an important
advantage to the smoothing method established so far.

The reliability of the model space smoothing method is confirmed
in the two cases: $^{58}$Ni($d$, $p n$) at 80~MeV
and the E1 transition of $^6$He.
In the former,
the new model space smoothing method is found to give the same
breakup $S$-matrix elements as the {\lq\lq}exact'' ones based on
the previous smoothing method in Ref.~\citen{Matsumoto}. \
In the latter, the E1 strength distribution obtained
by the model space smoothing method,
with the additional smearing procedure of Eq.~\eqref{eq:smear},
turns out to converge well as the model space is extended.

We apply the model space smoothing method to the $^6$He scattering
on $^{209}$Bi at 22.5 MeV, i.e., near the Coulomb barrier energy,
and calculate the angle-integrated energy spectrum of the
four-body breakup reaction.
Comparing the result of the four-body CDCC calculation with that of
the one-step calculation, we find that the one-step calculation is not
accurate at low incident energies.

Thus, the description of four-body breakup reactions with four-body CDCC
is accomplished by the model space smoothing method proposed in this
work. In a forthcoming paper, we will analyze the experimental data on
the angle-integrated energy spectrum $d\sigma/d\ve$~\cite{Aumann,Wang-MSU}
and the energy and angular correlations~\cite{Chulkov} among $^4$He
and the two neutrons in the breakup of $^6$He at lower and intermediate
incident energies.

\section*{Acknowledgements}

The authors would like to thank Prof. M.~Kamimura and Prof. Y.~Iseri for
helpful discussions.
This work has been supported in part by JSPS Research Fellowships
for Young Scientists and Grants-in-Aid for Scientific Research of
Monbukagakusyo of Japan.
One of the authors (T. M.) is grateful for the financial
assistance from the Special Postdoctoral Researchers Program of RIKEN.
The numerical calculations of this work were performed on the computing
system in Research Institute for Information Technology of Kyushu University.

\appendix

\section{Coordinate Transformation Between Rearrangement Channels}
\label{A1}

The coordinate transformation from
($\bm{r}_{b}, \bm{y}_{b}$) to ($\bm{r}_{c}, \bm{y}_{c}$) is defined by
\begin{align}
\begin{pmatrix}
\bm{r}_{b} \\ \bm{y}_{b}
\end{pmatrix}
=\begin{pmatrix}
  \g_{bc} &\d_{bc} \\
{\g'}_{bc}&{\d'}_{bc}
 \end{pmatrix}
\begin{pmatrix}
\bm{r}_{c} \\ \bm{y}_{c}
\end{pmatrix} .
\label{co-trans}
\end{align}
With this transformation,
the basis function $\varphi_{\alpha_b}^{(b)}(\bm{r}_b,\bm{y}_b)$
is rewritten as
\begin{align}
&\varphi^{(b)}_{\alpha_b}(\bm{r}_{c},\bm{y}_{c})
\notag\\
&=
\mathcal{N}_{i_{b}\ell_b}\mathcal{N}_{j_b\lambda_b}
e^{-\eta_{bc} r_{c}^2}e^{- 2\xi_{bc} \bm{r}_{c} \cdot \bm{y}_{c}}
e^{ -\zeta_{bc} y^2_{c}}
\sum_{\ell'_b,\lambda'_b,T}
{\langle \ell_b \lambda_b \Lambda_b | \ell'_b \lambda'_b T \Lambda_b
 \rangle}_{b \to c}
\notag\\
&\quad \times
r_{c}^{\ell_b+\lambda_b-T}y_{c}^{T}
\left[
\big[Y_{\ell'_b}(\Omega_{r_{c}}) \otimes
 Y_{\lambda'_b}(\Omega_{y_{c}})
 \big]_{\Lambda_b}
\otimes \big[ \eta_{1/2}^{(n_1)} \otimes \eta_{1/2}^{(n_2)}\big]_{S_b}
\right]_{Im} \ ,
\label{eq:03}
\end{align}
where $\alpha_b$ is the abbreviation of
$\{i_b,j_b,\ell_b,\lambda_b,\Lambda_b,S_b\}$ for channel $b$
and
\begin{align}
\eta_{bc} &=\nu_{i_b}\g_{bc}^2+\lambda_{j_b}{\g'}_{bc}^2,\\
\zeta_{bc}&=\nu_{i_b}\d_{bc}^2+\lambda_{j_b}{\d'}_{bc}^2,\\
\xi_{bc}  &=\nu_{i_b}\d_{bc}\g_{bc}
           +\lambda_{j_b}{\d'}_{bc}{\g'}_{bc}.
\end{align}
The coefficient
${\langle \ell_b \lambda_b \Lambda_b | \ell'_b \lambda'_b T \Lambda_b
 \rangle}_{b \to c}$ in Eq.~\eqref{eq:03}
is that of the Raynal transformation defined by
\begin{align}
&\sum_{\ell'_b,\lambda'_b,T}
 \langle \ell_b \lambda_b \Lambda_b | \ell'_b \lambda'_b T \Lambda_b
 \rangle_{b \to c}
 r_{c}^{\ell_b+\lambda_b-T}y_{c}^{T}
 \big[Y_{\ell'_b}(\Omega_{r_{c}}) \otimes
 Y_{\lambda'_b}(\Omega_{y_{c}})
 \big]_{\Lambda_b}
\notag\\
 & =
 (2\ell_b+1)(2\lambda_b+1)
 \sum_{\lambda=0}^{\ell_b}\sum_{\Lambda=0}^{\lambda_b}
 \sqrt{\frac{(2\ell_b)!}{(2\lambda)!(2\ell_b-2\lambda)!}}
 \sqrt{\frac{(2\lambda_b)!}{(2\Lambda)!(2\lambda_b-2\Lambda)!}}
 \notag\\
 &\times  {\g}_{bc}^{\ell_b-\lambda}{\g'}_{bc}^{\lambda_b-\Lambda}
 {\d}_{bc}^{\lambda}{\d'}_{bc}^{\Lambda}
 \sum_{\ell'_b \lambda'_b}
 \begin{Bmatrix}
  \ell_b-\lambda & \lambda_b-\Lambda & \ell'_b \\
  \lambda        & \Lambda     & \lambda'_b    \\
  \ell_b         & \lambda_b         & \Lambda_b
 \end{Bmatrix}
 \langle \ell_b-\lambda 0 \lambda_b-\Lambda 0 | \ell'_b 0 \rangle
 \langle \lambda        0 \Lambda     0 | \lambda'_b    0 \rangle
 \notag\\
 & \times
  r_{c}^{\ell_b+\lambda_b-\lambda+\Lambda}y_{c}^{\lambda+\Lambda}
 \big[Y_{\ell'_b}(\Omega_{r_{c}}) \otimes
 Y_{\lambda'_b}(\Omega_{y_{c}})
 \big]_{\Lambda_b} ,
\end{align}
where $T=\lambda+\Lambda$.

\section{Fourier Transform of $\hat{\Phi}_{nIm}(\bm{r},\bm{y})$}
\label{A2}

The Fourier transform $\tilde{\Phi}_{nIm}(\bm{k},\bm{K})$
of $\hat{\Phi}_{nIm}(\bm{r},\bm{y})$ has the form
\begin{align}
 \tilde{\Phi}_{nIm}(\bm{k},\bm{K})
 &=\frac{1}{(2\pi)^3}
 \iint d\bm{r}d\bm{y}e^{i\bm{k}\cdot\bm{r}}e^{i\bm{K}\cdot\bm{y}}
 \hat{\Phi}_{nIm}(\bm{r},\bm{y})
 \notag\\
 &=\frac{1}{(2\pi)^3}\sum_{c,\alpha}A_{\alpha}^{(c)nI}
 \tilde{\varphi}_{\alpha}^{(c)}(\bm{k}_c,\bm{K}_c) \
\end{align}
with
\begin{align}
&\tilde{\varphi}_{\alpha_c}^{(c)}(\bm{k}_c,\bm{K}_c)
=\mathcal{N}_{i_c\ell_c}
 \mathcal{N}_{j_c\lambda_c}
 \left(\frac{\pi}{\nu_{i_c}}\right)^{3/2}
 \left(\frac{\pi}{\lambda_{j_c}}\right)^{3/2}
 \left(\frac{i}{2\nu_{i_c}}\right)^{\ell_c}
 \left(\frac{i}{2\lambda_{j_c}}\right)^{\lambda_c}
 \notag\\
 &\times
 k_c^{\ell_c}e^{-{k_c^2}/{4\nu_{i_c}}}K_c^{\lambda_c}e^{-{K_c^2}/{4\lambda_{j_c}}}
 \left[
 \big[Y_{\ell_c}(\Omega_{k_c})\otimes Y_{\lambda_c}(\Omega_{K_c})
 \big]_{\Lambda_c}
 \otimes
 \big[\eta_{1/2}^{(n_1)}\otimes \eta_{1/2}^{(n_2)}\big]_{S_c}\right]_{Im}  .
\label{plane-wave-part}
\end{align}
Performing the coordinate transformation of Eq.~\eqref{co-trans} leads to
\begin{align}
&\tilde{\varphi}_{\alpha_b}^{(b)}(\bm{k}_c,\bm{K}_c)
\notag\\
&=\mathcal{N}_{i_b\ell_b}
 \mathcal{N}_{j_b\lambda_b}
 \left(\frac{\pi}{\nu_{i_b}}\right)^{3/2}
 \left(\frac{\pi}{\lambda_{j_b}}\right)^{3/2}
 \left(\frac{i}{2\nu_{i_b}}\right)^{\ell_b}
 \left(\frac{i}{2\lambda_{j_b}}\right)^{\lambda_b}
 \notag\\
 &\quad\times
 e^{-\eta'_{bc}k_c^2} e^{-2\xi'_{bc}\bm{k}_c\cdot\bm{K}_c} e^{-\zeta'_{bc}K_c^2}
 \sum_{\ell'_b,\lambda'_b,T}\langle\ell_b\lambda_b\Lambda_b
 |\ell'_b\lambda'_bT\Lambda_b \rangle_{b\to c}
 \notag\\
 &\quad \times
 k_c^{\ell_b+\lambda_b-T}K_c^{T}
 \left[\big[Y_{\ell'_b}(\Omega_{k_c})\otimes
 Y_{\lambda'_b}(\Omega_{K_c})\big]_{\Lambda_b}
 \otimes\big[\eta_{1/2}^{(n_1)}\otimes \eta_{1/2}^{(n_2)}\big]_{S_b}
 \right]_{Im} \
\label{plane-wave-part2}
\end{align}
with
\begin{align}
 \eta'_{bc} &=
 \frac{1}{4\nu_{i_b}}\bar{\g}^2_{bc}
 +\frac{1}{4\lambda_{j_b}}\bar{\g'}^2_{bc},
 \\
 \zeta'_{bc}&=
 \frac{1}{4\nu_{i_b}}\bar{\d}^2_{bc}
 +\frac{1}{4\lambda_{j_b}}\bar{\d'}^2_{bc},
 \\
 \xi'_{bc} &=\frac{1}{4\nu_{i_b}}\bar{\d}_{bc}\bar{\g}_{bc}
 +\frac{1}{4\lambda_{j_b}}\bar{\d'}_{bc}\bar{\g'}_{bc},
\end{align}
where
\begin{equation}
 \bar{\g}_{bc}=\g_{cb}, \  \bar{\d}_{bc}=\g'_{cb}, \
\bar{\g'}_{bc}=\d_{cb}, \ \bar{\d'}_{bc}=\d'_{cb}.
\end{equation}

\section{Matrix Elements of $G_{ij}$}
\label{A3}

In $G_{ij}$ the six-fold integration is reduced to a
single one:
\begin{align}
 G_{ij}
 & =\sum_{a=1}^{3}\sum_{b=1}^{3}\sum_{\alpha_a}\sum_{\alpha_b}
 A_{\alpha_a}^{(a)iI}A_{\alpha_b}^{(b)jI}\tilde{G}_{\alpha_a\alpha_b},
\label{eq:m-green0}
\end{align}
where $\alpha_c=\{i_c, j_c, \ell_c, \lambda_c, \Lambda_c, S_c$\}
for channel $c$ ($=$1--3) and $\tilde{G}_{\alpha_a\alpha_b}$ is
given by
\begin{align}
 &\tilde{G}_{\alpha_a\alpha_b} \notag\\
 &= \d_{\Lambda_a\Lambda_b}\d_{SS'}
 \frac{(-)^{\Lambda_b}}{(2\pi)^6}
 \left(\frac{\pi}{\nu_{i_a}} \frac{\pi}{\lambda_{j_a} }
 \frac{\pi}{\nu_{i_b} }\frac{\pi}{\lambda_{j_b} }\right)^{3/2}
 \left(\frac{-i}{2\nu_{i_a} }\right)^{\ell_a}
 \left(\frac{-i}{2\lambda_{j_a} }\right)^{\lambda_a}
 \left(\frac{i}{2\nu_{i_b} }\right)^{\ell_b}
 \left(\frac{i}{2\lambda_{j_b} }\right)^{\lambda_b}
 \notag\\
 &\times
 \sum_{\ell'_a\lambda'_aT}
 \langle \ell_a \lambda_a \Lambda_a | \ell'_a\lambda'_aT \Lambda_a
 \rangle_{a \to b}
 \hat{\ell}_a'\hat{\lambda}_a'\hat{\ell}_b\hat{\lambda}_b
 \sum_{\kappa}
 W(\lambda_a', \lambda_b,\ell_a', \ell_b;  \kappa \Lambda_b )
 \notag\\
 &\times
 \langle \ell_a' 0 \ell_b 0 | \kappa 0\rangle
 \langle \lambda_a'    0 \lambda_b    0 | \kappa 0\rangle
 \frac{\pi(2m+2\kappa+1)!!}{2^{m+\kappa+n+4}}
 \sum_{k=0}^{m}\frac{(2k+2\kappa+2n+1)!!}{(2k+2\kappa+1)!!}
 \binom{m}{k} \notag\\
&\times
  (-i)
 \int_0^{\infty}  d\tau e^{i\tau \ve}
 \frac{\left(\eta'+i\tau\tilde{\mu}_r\right)^{n-m}
 (\xi')^{2k+\kappa}}
 {\left((\zeta'+i\tau\tilde{\mu}_y)
 (\eta'+i\tau\tilde{\mu}_r)
 -{(\xi')^2}
 \right)^{k+\kappa+n+3/2}}
 \label{eq:m-green}
\end{align}
with
\begin{align*}
 \tilde{\mu}_r&=\hbar^2k_b^2/(2\mu_{r_b}), &
\tilde{\mu}_y&=\hbar^2K_b^2/(2\mu_{y_b}), \\
\eta' &= \eta'_{ab} +{1}/{(4\nu_{i_b})} , &
\zeta'&= \zeta'_{ab} +{1}/{(4\lambda_{j_b})},&\
\xi' = \xi'_{ab},  \\
2n&=\ell_a+\ell_b+\lambda_a-T-\kappa,  &
2m&=\lambda_b+T-\kappa,
\end{align*}
and $\hat{\ell}=\sqrt{2\ell+1}$ and
$W(a, b, c, d ;e, f)$ is the Racah coefficient.
The use of Eqs.~\eqref{plane-wave-part2} and \eqref{eq:m-green}
greatly simplifies numerical calculations.


\end{document}